\documentstyle[12pt]{article}
\date{\small}
\title{\bf Classical and Quantum Cosmology of an Accelerating Model Universe with Compactification of Extra Dimensions}
\author{F. Darabi\thanks{e-mail:
f.darabi@azaruniv.edu}\\{\small Department of Physics, Azarbaijan
University of Tarbiat Moallem, 53714-161, Tabriz, Iran .}}
\begin{document}
\maketitle
\begin{abstract}
We study a $(4+D)$-dimensional Kaluza-Klein cosmology with a
Robertson-Walker type metric having two scale factors $a$ and $R$,
corresponding to $D$-dimensional internal space and 4-dimensional
universe, respectively. By introducing an exotic matter in the
form of perfect fluid with an special equation of state, as the
space-time part of the higher dimensional energy-momentum tensor,
a four dimensional effective decaying cosmological term appears as
$\lambda \sim R^{-m}$ with $0 \leq m\leq 2$, playing the role of
an evolving dark energy in the universe. By taking $m=2$, which
has some interesting implications in reconciling observations with
inflationary models and is consistent with quantum tunneling, the
resulting Einstein's field equations yield the exponential
solutions for the scale factors $a$ and $R$. These exponential
behaviors may account for the dynamical compactification of extra
dimensions and the accelerating expansion of the 4-dimensional
universe in terms of Hubble parameter, $H$. The acceleration of
the universe may be explained by the negative pressure of the
exotic matter. It is shown that the rate of compactification of
higher dimensions as well as expansion of 4-dimensional universe
depends on the dimension, $D$. We then obtain the corresponding
Wheeler-DeWitt equation and find the general exact solutions in
$D$-dimensions. A good correspondence between the solutions of
classical Einstein's equations and the solutions of quantum
Wheeler-DeWitt equation in any dimension, $D$, is obtained based
on Hartle's point of view concerning the classical limits of
quantum cosmology.
\end{abstract}

\newpage
\section{Introduction}
Cosmological models with a cosmological term $\Lambda$ are
currently serious candidates to describe the dynamics of our four
dimensional universe. The history of cosmological term dates back
to Einstein, and its original role was to allow static homogeneous
solutions to Einstein's equations in the presence of matter which
turned out to be unnecessary when the expansion of the universe
was discovered . However, particle physicists then realized that
the non-vanishing cosmological constant can be interpreted as a
measure of the energy density of the vacuum which turned out to be
the sum of a number of apparently disjoint contributions of
quantum fields. In fact, a dynamical characteristic for the vacuum
energy density (cosmological term) was attributed by quantum field
theorists since the developments in particle physics and
inflationary scenarios. According to modern quantum field theory,
the structure of a vacuum turns out to be interrelated with some
spontaneous symmetry-breaking effects through the condensation of
quantum (scalar) fields. This phenomenon gives rise to a
non-vanishing vacuum energy density of the form $<T_{\mu
\nu}>=-<\rho>g_{\mu \nu}$. Therefore, the observed (or effective)
cosmological term receives an extra contribution from $<T_{\mu
\nu}>$ as follows:
$$
\Lambda=\lambda+8\pi G<\rho>,
$$
where $\lambda$ is the bare cosmological constant and $G$ is the
gravitational constant. From quantum field theory we may expect
$<\rho> \approx M_{Pl}^4 \approx 2\times 10^{71} GeV^4$ ($M_{Pl}$
is the Planck mass), or another energy scale related to some
spontaneous symmetry breaking effect such as $M_{SUSY}^4$ or
$M_{Weak}^4$. Therefore, the bare cosmological constant receives
potential contributions from these mass scales resulting in a
large effective cosmological term. However, the experimental upper
bound on the present value of the cosmological term, $\Lambda$,
provided by measurements of the Hubble constant, $H$, reads
numerically as
$$
\frac{|\Lambda|}{8\pi G} \leq 10^{-29} g/cm^3 \approx 10^{-47}
GeV^4,
$$
which is too far from the expectation of quantum field theory .
The question of why the observed vacuum energy is so small in
comparison to the scales of particle physics is known as the
cosmological constant problem. It is generally thought to be
easier to imagine an unknown mechanism which would set $\Lambda$
exactly to zero than one which would suppress it by just the right
amount to yield an observationally tiny cosmological constant. If
$\Lambda$ is a dynamical variable (or vacuum parameter), then it
is natural to suppose that in an expanding universe the
cosmological term relaxes to the present tiny value by some
relaxation mechanism which may be provided by a time-varying
vacuum with a rolling scalar field \cite{1'}.

There are still other possibilities to be advocated. In recent
years, several attempts in these directions have been done, in the
context of quantum cosmology \cite{2'}. One plausible explanation
for a tiny cosmological term is to suppose that $\Lambda$ is
dynamically evolving and not constant, i.e., $\Lambda \propto
R^{-m}$, where $R$ is the scale factor of the universe and $m$ is
a parameter. So, as the universe expands from its small size in
the early universe, the initially large effective cosmological
term evolves and reduces to its present small value\cite{3'}.

The study of $\Lambda$-decaying cosmological models has recently
been the subject of particular interest both from classical and
quantum aspects. The $\Lambda$ decaying models may serve as
potential candidates to solve this problem by decaying the large
value of the cosmological constant $\Lambda$ to its present
observed value.

Also, there are strong (astronomical) observational motivations
for considering cosmological models in which $\Lambda$ is
dynamically decreasing as $\Lambda \propto R^{-m}$. Some models
assume {\em a priori} a fixed value for the parameter $m$. The
case $m=2$, corresponding to the cosmic string matter, has mostly
been taken based on dimensional considerations by some authors
\cite{1}. The case $m \approx 4$ which resembles the ordinary
radiation has also been considered by some other authors \cite{2}.
A third group of authors have also studied the case $m=3$
corresponding to the ordinary matter \cite{3}. There are also some
other models in which the value of $m$ is not fixed {\em a priori}
and the numerical bounds on the value of $m$ is estimated by
observational data or obtained by calculation of the quantum
tunnelling rate \cite{4}. Other aspects of $\Lambda$-decaying
models have also been discussed with no specific numerical bounds
on $m$ \cite{5}. It is clear that the functional dependence
$\Lambda \propto R^{-m}$ is phenomenological and does not result
from the first principles of particle physics. However, for some
domain for example, $0\leq m <3$, the decaying law $\Lambda
\propto R^{-m}$ deserves further investigation. One important
reason is that the age of the universe, in these models, is always
larger than the age obtained in the standard Einstein-de Sitter
cosmology, or the one we get in an open universe. Therefore, if we
are interested in solving the age problem, the decaying $\Lambda$
term appears to be a good candidate. In fact, according to the
ansatz $\Lambda \propto R^{-m}$, one may suppose the natural value
$<\rho> \approx M_{Pl}^4 $ to be the value of $\Lambda$ at the
Planck time when $R$ was of the order of the Planck length.
Theoretically this ansatz does not directly solve the cosmological
constant problem, but it relates this problem to the age problem
of why our universe is so old and have a radius $R$ much larger
than the Planck length. In other words, this ansatz reduces two
above problems to one problem of ``Why our universe could have
escaped the death at the Planck time'', which seems to be the most
natural fate of a baby-universe in quantum cosmology? One may
assume that the value of $\Lambda$ in the early universe might
have been much bigger than its present value and large enough to
drive some symmetry breakings which might have occurred in the
early universe.

On the other hand, the idea that our 4-dimensional universe might
have emerged from a higher dimensional space-time is now receiving
much attention \cite{6} where the compactification of higher
dimensions plays a key role. However, the question of how and why
this compactification occurs remains as an open problem. From
string theory we know that the compactification may take place
provided that the higher dimensional manifold admits special
properties, namely if the geometry of the manifold allows, for
example, the existence of suitable Killing vectors. However, it is
difficult to understand why such manifolds are preferred and
whether other possible mechanisms for compactification do exist.
In cosmology, on the other hand, different kinds of
compactifications could be considered. For example, in an
approach, called {\it dynamical compactification} , the extra
dimensions evolve in time towards very small sizes and the
extra-dimensional universe reduces to an effective
four-dimensional one. This type of compactification was considered
in the context of Modern Kaluza-Klein theories \cite{7}. It is
then a natural question that how an effective four dimensional
universe evolve in time and whether the resulting cosmology is
similar to the standard Friedmann-Robertson-Walker four
dimensional universe without extra dimensions.

Meanwhile, the recent distance measurements of type Ia {\it
supernova} suggest strongly an accelerating universe \cite{A}.
This accelerating expansion is generally believed to be driven by
an energy source called {\it dark energy} which provides negative
pressure, such as a positive cosmological constant \cite{q}, or a
slowly evolving real scalar field called {\it quintessence}
\cite{!}. Moreover, the basic conclusion from all previous
observations that $\sim$ 70 percent of the energy density of the
universe is in a dark energy sector, has been confirmed after the
recent WMAP \cite{!!}.

To model a universe based on these considerations one may start
from a fundamental theory including both gravity and standard
model of particle physics. In this regards, it is interesting to
begin with ten or eleven-dimensional space-time of
superstring/M-theory, in which case one needs a compactification
of ten or eleven-dimensional supergravity theory where an
effective 4-dimensional cosmology undergoes acceleration. However,
it has been known for some time that it is difficult to derive
such a cosmology and has been considered that there is a no-go
theorem that excludes such a possibility, if one takes the
internal space to be time-independent and compact without boundary
\cite{!!!}. However, it has recently been shown that one may avoid
this no-go theorem by giving up the condition of time-independence
of the internal space; and a solution of the vacuum Einstein
equations with compact hyperbolic internal space has been proposed
based on this model \cite{&&}. Similar accelerating cosmologies
can also be obtained for SM2 and SD2 branes , not only for
hyperbolic but also for flat internal space \cite{&&&}.

On the other hand, from cosmological point of view, it is not so
difficult to find cosmological models in which the 4-dimensional
universe undergoes an accelerating expansion and the internal
space contracts with time, exhibiting the {\it dynamical
compactification} \cite{7}, \cite{8}, \cite{9}.

In \cite{9}, for instance, it is shown that using a more general
metric, as compared to Ref.\cite{7}, and introducing matter
without specifying its nature, the size of compact space evolves
as an inverse power of the radius of the universe. The
Friedmann-Robertson-Walker equations of the standard
four-dimensional cosmology is obtained using an effective pressure
expressed in terms of the components of the higher dimensional
energy-momentum tensor, and the negative value of this pressure
may explain the acceleration of our present universe.

To the author's knowledge the question of $\Lambda$-decaying
cosmological model has not received much attention in higher
dimensional Kaluza-Klein cosmologies. Moreover, the exotic matter
has not been considered as an alternative candidate to produce the
acceleration of the universe. The purpose of the present chapter
is to study a $(4+D)$-dimensional Kaluza-Klein cosmology ,with an
extended Robertson-Walker type metric, in this context
\cite{paper}. As we are concerned with cosmological solutions,
which are intrinsically time dependent, we may suppose that the
internal space is also time dependent. It is shown that by taking
this higher dimensional metric and introducing a 4-dimensional
exotic matter, a decaying cosmological term $\Lambda \sim R^{-m}$
with $0\leq m \leq 2$ is appeared as a type of dark energy, and
for the case $m=2$ the resulting field equations yield the
exponential solutions for the scale factors of the
four-dimensional universe and the internal space. These solutions
may account for the accelerating universe and dynamical
compactification of extra dimensions, driven by the negative
pressure of the exotic matter \footnote{A similar work \cite{@@}
has already been done in which the same extended FRW metric was
chosen with a radiation fluid occupying all the extended
space-time. They found an inflation for 3-dimensions and a
contraction for the $D$ remaining spatial dimensions. }. It should
be noted, however, that the solutions in principle describe
typical inflation rather than the recently observed acceleration
of the universe which is known to take place in an ordinary matter
dominated universe. Nevertheless, regarding the fact that about 70
percent of the total energy density of the universe is of dark
energy type with negative pressure, we may approximate the matter
content of the universe with almost dark energy and consider the
present model as a rather simplified model of a real accelerating
universe.

The quantum cosmology of this model is also studied by obtaining
the Wheeler-DeWitt equation and finding its general exact
solutions. It is then shown that a good correspondence exists
between the classical and quantum cosmological solutions, based on
the interpretation of Hartle of the classical limits of quantum
cosmology.

The chapter is organized as follows: In section {\bf 2}, we
introduce the classical cosmology model by taking a higher
dimensional Robertson-Walker type metric and a higher dimensional
matter whose non-zero part is a four-dimensional exotic matter. In
section {\bf 3}, we obtain the Einstein equations for the two
scale factors. In section {\bf 4}, we solve the Einstein equations
and obtain the solutions. In section {\bf 5}, we study the
corresponding quantum cosmology and derive the Wheeler-DeWitt
equation. In section {\bf 6}, the exact solutions of the
Wheeler-DeWitt equation is obtained. Finally, in section {\bf 7},
we show a good correspondence between the classical and quantum
cosmology. The chapter is ended with concluding remarks.

\section{Classical cosmology}

To begin with, we study the metric considered in \cite{10} in
which the space-time is assumed to be of Robertson-Walker type
having a (3+1)-dimensional space-time part and an internal space
with dimension $D$. We adopt a real chart $\{t, r^{i}, \rho^{a}\}$
with $t$, $r^{i}$, and $\rho^{a}$ denoting the time, space
coordinates and internal space dimensions, respectively. We,
therefore, take\footnote{There is a little difference between this
metric and that of \cite{10}, in that here the lapse function is
generally considered as $N(t)$ instead of taking $N=1$.}
\begin{equation}
ds^2=-N^2(t) dt^2+R^2(t)\frac{dr^i
dr^i}{(1+\frac{kr^2}{4})^2}+a^2(t) \frac{d\rho^a
d\rho^a}{(1+k^\prime \rho^2)}, \label{1}
\end{equation}
where $N(t)$ is the lapse function, $R(t)$ and $a(t)$ are the
scale factor of the universe and the radius of internal space,
respectively; $r^2 \equiv r^i r^i (i=1, 2, 3), \rho^2\equiv \rho^a
\rho^a (a=1, ... D)$, and $k, k^\prime=0, \pm1$, reflecting flat,
open or closed type of four-dimensional universe and
$D$-dimensional space. We assume the internal space to be flat
with compact topology $S^D$, which means $k^\prime =0$. This
assumption is motivated by the possibility of the compact spaces
to be flat or hyperbolic in ``{\it accelerating cosmologies from
compactification}'' scenarios, as discussed in Introduction.

The form of energy-momentum tensor is dictated by Einstein's
equations and by the symmetries of the metric (\ref{1}).
Therefore, we may assume
\begin{equation}
T_{AB}=( -\rho, p, p, p, p_{_D}, p_{_D}, ..., p_{_D} ), \label{3}
\end{equation}
where $A$ and $B$ run over both the space-time coordinates and the
internal space dimensions. Now, we examine the case for which the
pressure along all the extra dimensions vanishes, namely
$p_{_D}=0$. In so doing, we are motivated by the {\it brane world}
scenarios where the matter is to be confined to the 4-dimensional
universe, so that all components of $T_{AB}$ is set to zero but
the space-time components \cite{!&} and it means no matter escapes
through the extra dimensions.

We assume the energy-momentum tensor $T_{\mu \nu}$ of space-time
to be an exotic $\chi$ fluid with the equation of state
\begin{equation}
p_\chi =(\frac{m}{3}-1)\rho_\chi, \label{4}
\end{equation}
where $p_\chi$ and $\rho_\chi$ are the pressure and density of the
fluid, respectively and the parameter $m$ is restricted to the
range $0\leq m \leq 2$ \cite{12}. It is worth noting that the
equation of state (\ref{4}) with $0\leq m \leq 2$ resembles a
universe with negative pressure matter, violating the strong
energy condition \cite{15} and this violation is required for a
universe to be accelerated \cite{&&}\footnote{Given Einstein
equations, this condition on the energy-momentum tensor implies a
condition on Ricci tensor as $R_{00}\geq 0$.}.

Using standard techniques we obtain the scalar curvature
corresponding to the metric (\ref{1})
$$
{\cal
R}=\frac{-6RaN\ddot{R}+6Ra\dot{N}\dot{R}-2R^2\ddot{a}N+2R^2\dot{N}\dot{a}-2aN^3k+aN^3k^2
r^2-6aN\dot{R}^2 -6R\dot{R}\dot{a}N}{R^2N^3a},
$$
and then substitute it into the dimensionally extended
Einstein-Hilbert action ( without higher dimensional cosmological
term) plus a matter term indicating the above mentioned exotic
fluid. This leads to the effective Lagrangian \footnote{We take
the Planck units, $G=c=\hbar=1$}
\begin{equation}
L=\frac{1}{2N}Ra^D{\dot{R}}^2+\frac{D(D-1)}{12N}R^3a^{D-2}{\dot{a}}^2+
\frac{D}{2N}R^2a^{D-1}\dot{R}\dot{a}-\frac{1}{2}kNRa^D+\frac{1}{6}N\rho_{\chi}
R^3a^D, \label{5}
\end{equation}
where a dot represents differentiation with respect to $t$. We now
take a closed $(k=1)$ universe. Although the flat universe $(k=0)$
is almost favored by observations, we will show an equivalence
between $(k=1)$ and $(k=0)$ universes. One may obtain the
continuity equation by using the contracted Bianchi identity in
(4+$D$) dimensions, namely
$$
\nabla_M G^{MN}=\nabla_M T^{MN}=0,
$$
together with the assumption that the matter is confined to
(3+1)-dimensional space-time as
$$
T_{a b}=T_{\mu a}=0,
$$
which gives rise to
$$
\nabla_\mu T^{\mu \nu}=0,
$$
or
\begin{equation}
\dot{\rho}_\chi R+3(p_\chi+\rho_\chi)\dot{R}=0. \label{6}
\end{equation}
It is easily shown that substituting the equation of state
(\ref{4}) into the continuity equation (\ref{6}) leads to the
following behavior of the energy density in a closed ($k=1$)
Friedmann-Robertson-Walker universe \cite{12}
\begin{equation}
\rho_\chi(R)=\rho_\chi(R_0)\left(\frac{R_0}{R}\right)^m_,
\label{7}
\end{equation}
where $R_0$ is the value of the scale factor at an arbitrary
reference time $t_0$.

Now, if we believe that the cosmological term plays the role of
vacuum energy density, we may define the cosmological term
\cite{4}
\begin{equation}
\Lambda\equiv\rho_\chi(R), \label{8}
\end{equation}
which leads to
\begin{equation}
L=\frac{1}{2N}Ra^D{\dot{R}}^2+\frac{D(D-1)}{12N}R^3a^{D-2}{\dot{a}}^2+
\frac{D}{2N}R^2a^{D-1}\dot{R}\dot{a}-\frac{1}{2}NR
a^D+\frac{1}{6}N\Lambda R^3a^D, \label{9}
\end{equation}
where the cosmological term is now decaying with the scale
factor$R$ as
\begin{equation}
\Lambda(R)=\Lambda(R_0)\left(\frac{R_0}{R}\right)^m_. \label{10}
\end{equation}
Note that $\Lambda$ is now playing the role of an evolving dark
energy \cite{*} in 4-dimensions, because we did not consider
explicitly a $(4+D)$ dimensional cosmological term in the action,
and $\Lambda$ appears merely due to the specific choice of the
equation of state (\ref{4}) for the exotic matter. The decaying
$\Lambda$ term may also explain the smallness of the present value
of the cosmological constant since as the universe evolves from
its small to large sizes the large initial value of $\Lambda$
decays to small values. This phenomenon may somehow alleviate the
cosmological constant problem.

Of particular interest, to us, among the different values of $m$
is $m=2$ which has some interesting implications in reconciling
observations with inflationary models \cite{13}, and is consistent
with quantum tunnelling \cite{4}.

\section{Einstein equations}

We take $m=2$ and set the initial values of $R_0$ and
$\Lambda(R_0)$ as
\begin{equation}
\Lambda(R_0)R^2_0=3 \:\:\:, \:\:\:\Lambda(R)=\frac{3}{R^2},
\label{11}
\end{equation}
leading to a positive cosmological
term which, according to (\ref{8}), guarantees the weak energy
condition $\rho_\chi >0$.

The lapse function $N(t)$, in principle, is also an arbitrary
function of time due to the fact that Einstein's general
relativity is a reparametrization invariant theory. We, therefore,
take the gauge
\begin{equation}
N(t)=R^3(t) a^D(t). \label{12}
\end{equation}
Now, the Lagrangian becomes
\begin{equation}
L=\frac{1}{2}\frac{\dot{R}^2}{R^2}+\frac{D(D-1)}{12}\frac{\dot{a}^2}{a^2}+\frac{D}{2}\frac{\dot{R}\dot{a}}{Ra},
\label{13}
\end{equation}
where Eq.(\ref{11}) has been used. It is seen that the parameters
$k$  and $\Lambda$ are effectively removed from the Lagrangian and
this implies that although $k$ and $\Lambda$ are not zero in this
model the corresponding 4-dimensional universe is equivalent to a
flat universe with a zero cosmological term. In other words, we do
not distinguish between our familiar 4-dimensional universe, which
seems to be flat and without any exotic fluid, and a closed universe
filled with an exotic fluid.\\
We now define the new variables
\begin{equation}
X=\log{R}\:\:\:,\:\:\:Y=\log{a}. \label{14}
\end{equation}
The lagrangian (\ref{13}) is written as
\begin{equation}
L=\frac{1}{2}\dot{X}^2+\frac{D(D-1)}{12}\dot{Y}^2+\frac{D}{2}\dot{X}\dot{Y}.
\label{15}
\end{equation}
The equations of motion are obtained
\begin{equation}
\ddot{X}+\frac{D}{2}\ddot{Y}=0, \label{16}
\end{equation}
\begin{equation}
\ddot{X}+\frac{D-1}{3}\ddot{Y}=0. \label{17}
\end{equation}
Combining the equations (\ref{16}) and (\ref{17}) we obtain
\begin{equation}
\ddot{X}=0, \label{18}
\end{equation}
\begin{equation}
\ddot{Y}=0. \label{19}
\end{equation}

\section{Solutions of Einstein equations}

The solutions for $X$ and $Y$ in Eqs. (\ref{18}) and (\ref{19})
are obtained
\begin{equation}
X={\cal A}t+ \gamma,
\end{equation}
\begin{equation}
Y={\cal B}t+ \delta,
\end{equation}
and the solutions for $R(t)$ and $a(t)$ are then as follows
\begin{equation}
R(t)=Ae^{\alpha t}, \label{20}
\end{equation}
\begin{equation}
a(t)=Be^{\beta t}, \label{21}
\end{equation}
where the constants ``${\cal A}$, ${\cal B}$, $\gamma$ and
$\delta$'' or ``$A$, $B$, $\alpha$ and $\beta$'' should be
obtained, in principle, in terms of the initial conditions. It is
a reasonable assumption that the size of all spatial dimensions be
the same at $t=0$. Moreover, it may be assumed that this size
would be the Planck size $l_p$ in accordance with quantum
cosmological considerations. Therefore, we take $R(0)=a(0)={\it
l_P}$ so that $A=B={\it l_p}$, and
\begin{equation}
R(t)={\it l_p}e^{\alpha t}, \label{22}
\end{equation}
\begin{equation}
a(t)={\ l_p}e^{\beta t}. \label{23}
\end{equation}
It is important to note that the constants $\alpha, \beta$  are
not independent, and a relation may be obtained between them. This
is done by imposing the zero energy condition $H=0$ which is the
well-known result in cosmology due to the existence of arbitrary
laps function $N(t)$ in the theory. The Hamiltonian constraint is
obtained through the Legender transformation of the Lagrangian
(\ref{15})
\begin{equation}
H=\frac{1}{2}\dot{X}^2+\frac{D(D-1)}{12}\dot{Y}^2+\frac{D}{2}\dot{X}\dot{Y}=0,
\label{24}
\end{equation}
which is written in terms of  $\alpha$ and $\beta$ as
\begin{equation}
H=\frac{1}{2}\alpha^2+\frac{D(D-1)}{12}\beta^2+\frac{D}{2}\alpha\beta=0.
\label{25}
\end{equation}
This constraint is satisfied only for $\alpha\le 0, \beta\ge 0$ or
$\alpha\ge 0, \beta\le 0$.

For $D \neq 1$, the case $\alpha=0$ or $\beta=0$ gives rise to
time independent scale factors, namely $R=a=l_P$, which is not
physically viable since we know, at least based on observations,
the scale factor of the universe is time dependent. We, therefore,
choose $\alpha>0 , \beta<0$ so that the universe and the internal
space would expand and contract, respectively, in accordance with
the present observations.

For the case $D=1$, we find
\begin{equation}
\left \{ \begin{array}{ll}
\beta=\mbox{arbitrary} \\
\alpha=0
\end{array}\right.
\:\: \mbox{or} \:\:\:\: \alpha=-\beta. \label{26}
\end{equation}
The former is not physically viable, since it predicts no time
evolution for the universe. The latter, however, may predict
exponential expansion for $R(t)$, and exponential contraction for
$a(t)$, both with the same exponent $\alpha>0$.

For the general case $D>1$, we find
\begin{equation}
\alpha_\pm=\frac{D\beta}{2}\left[-1\pm
\sqrt{1-\frac{2}{3}(1-\frac{1}{D})}\right], \label{27}
\end{equation}
which gives two positive values for $\alpha$ indicating two
possible expanding universes provided $\beta<0$ which indicates
the compactification of extra dimensions. Moreover, the values of
$\alpha_\pm$, for a given negative value of $\beta$, become larger
for higher dimensions. Therefore, the universe expands more
rapidly in both possibilities. On the contrary, for a given
positive value of $\alpha$, indicating an expanding universe, the
parameter $\beta$ may take two negative values
\begin{equation}
\beta_{\pm}=\frac{2 \alpha}{D}\left[-1\pm
\sqrt{1-\frac{2}{3}(1-\frac{1}{D})}\right]^{-1},
\end{equation}
indicating two ways of compactification. Moreover, they become
smaller for higher dimensions, exhibiting lower rates of
compactification.

To find the constants $\alpha, \beta$ we first obtain the Hubble
parameter for $R(t)$
\begin{equation}
H=\frac{\dot{R}}{R}=\alpha, \label{30}
\end{equation}
by which the constant $\alpha$ is fixed. The observed positive
value of $H$ will then justify our previous assumption,
$\alpha>0$. We may, therefore, write the solutions (\ref{22}) and
(\ref{23}) in terms of the Hubble parameter $H$ as
\begin{equation}
R(t)=l_p e^{Ht}, \label{30'}
\end{equation}
\begin{equation}
a(t)=l_p e^{-Ht}, \label{30''}
\end{equation}
for $D=1$, and
\begin{equation}
R(t)=l_p e^{Ht}, \label{31}
\end{equation}
\begin{equation}
a(t)_\pm=l_p e^{\frac{2Ht}{D}\left[-1\pm
\sqrt{1-\frac{2}{3}(1-\frac{1}{D})}\right]^{-1}}, \label{32}
\end{equation}
and
\begin{equation}
R_{\pm}(t)=l_p e^{\frac{D \beta t}{2}\left[-1\pm
\sqrt{1-\frac{2}{3}(1-\frac{1}{D})}\right]}, \label{0}
\end{equation}
\begin{equation}
a(t)=l_p e^{\beta t}. \label{00}
\end{equation}
for $D>1$.

For a given $H>0$, it is seen that the solution corresponding to
$D=1$ may predict an accelerating (de Sitter) universe and a
contracting internal space with exactly the same rates. For $D>1$,
in Eqs.(\ref{31}) and (\ref{32}), for a given $H>0$ in the
exponent of $R(t)$ the exponent in $a(t)$ takes two negative
values and becomes smaller for higher dimensions. This means that
while the 4-dimensional (de Sitter) universe is expanding by the
rate $H$, the higher dimensions may be compactified in two
possible ways with different rates of compactification as a
function of dimension, $D$. In Eqs.(\ref{0}) and (\ref{00}), on
the other hand, for a given $\beta<0$ the exponent in $R(t)$ takes
two positive values which become larger for higher dimensions.
This also means that while the extra dimensions contract by the
rate $\beta$, the universe may be expanded in two possible ways
with different expansion rates as a function of $D$.

It is easy to show that the Lagrangian (\ref{15}) ( or the
equations of motion ) is invariant under the simultaneous
transformation
\begin{equation}
R\rightarrow R^{-1}\:\:\:\:,\:\:\:\: a\rightarrow a^{-1},
\label{28}
\end{equation}
which is consistent with the time reversal $t\rightarrow -t$.
Therefore, four different phases of ``{\it
expansion-contraction}'' for $R(t)$ and $a(t)$ are distinguished,
Eqs.(\ref{31}) - (\ref{00}). One may prefer the ``{\it expanding
$R(t)$ - contracting $a(t)$}'' phase to ``{\it expanding $a(t)$ -
contracting $R(t)$}'' one, considering the present status of the
4D universe \footnote{For the special case $D=3$, both the
Lagrangian (\ref{15}) and the Hamiltonian constraint (\ref{24})
are invariant under the transformation
$$
a\rightarrow R \:\:\:\:,\:\:\:\: R \rightarrow a.
$$
Therefore, we have a dynamical symmetry between $R$ and $a$,
namely
$$
a \leftrightarrow R.
$$
In this case there is no real line of demarcation between $a$ and
$R$ to single out one of them as the real scale factor of the
universe. This is because the internal space is flat $k'=0$ and
according to (\ref{13}) one may assume the 4D universe with $k,
\Lambda\neq0$ to be equivalent to the one in which $k=\Lambda=0$.
Therefore, both have the same topology $S^3$.}.

The deceleration parameter $q$ for the scale factor $R$ is
obtained
\begin{equation}
q=-\frac{\ddot{R}R}{\dot{R}^2}=-1. \label{29}
\end{equation}
Observational evidences not only do not rule out the negative
deceleration parameter but also puts the limits on the present
value of $q$ as $-1\leq q < 0$ \cite{A}. Therefore, this negative
value seems to favor a cosmic acceleration in the expansion of the
universe.

In the expansion phase of the closed ($k=1$) universe the
cosmological term $\Lambda$ decays exponentially with time $t$ as
\begin{equation}
\Lambda(t)=3l_p^{-2}e^{-2Ht}, \label{33}
\end{equation}
whereas in the contraction phase ($t \rightarrow -t$) it grows
exponentially to large values so that at $t=0$ it becomes
extremely large, of the order of $M_p^2$. This huge value of
$\Lambda$ may be extinguished rapidly by assuming a sufficiently
large Hubble parameter $H$, consistent with the present
observations, to alleviate the cosmological constant problem.

\section{Quantum cosmology}

An appropriate quantum mechanical description of the universe is
likely to be afforded by quantum cosmology which was introduced
and developed by DeWitt \cite{4'}. In quantum cosmology the
universe, as a whole, is treated quantum mechanically and is
described by a single wave function, $\Psi(h_{ij}, \phi)$, defined
on a manifold ({\it superapace}) of all possible three geometries
and all matter field configurations.  The wave function
$\Psi(h_{ij}, \phi)$ has no explicit time dependence due to the
fact that there is no a real time parameter external to the
universe. Therefore, there is no Schr$\ddot{o}$dinger wave
equation but the operator version of the Hamiltonian constraint of
the Dirac canonical quantization procedure \cite{Dirac}, namely
vanishing of the variation of the Einstein-Hilbert action $S$ with
respect to the arbitrary lapse function $N$
$$
H=\frac{\delta S}{\delta N}=0,
$$
which is written
$$
\hat{H}\Psi(h_{ij}, \phi)=0.
$$
This equation is known as the Wheeler-DeWitt (WDW) equation. The
goal of quantum cosmology by solving the WDW equation is to
understand the origin and evolution of the universe, quantum
mechanically. As a differential equation, the WDW equation has an
infinite number of solutions. To get a unique viable solution , we
should also respect the question of boundary condition in quantum
cosmology which is of prime importance in obtaining the relevant
solutions for the WDW equation.

In principle, it is very difficult to solve the WDW equation in
the {\it superspace} due to the large number of degrees of
freedom. In practice, one has to {\it freeze out} of all but a
finite number of degrees of freedom of the gravitational and
matter fields. This procedure is known as quantization in {\it
minisuperspace}, and will be used in the following discussion.

The minisuperspace in our model is two-dimensional with
gravitational variables $X$ and $Y$. To obtain the Wheeler-DeWitt
equation, in this minisuperspace, we start with the Lagrangian
(\ref{15}). The conjugate momenta corresponding to $X$ and $Y$ are
obtained
\begin{equation}
P_X=\frac{\partial L}{\partial
\dot{X}}=\dot{X}+\frac{D}{2} \dot{Y},
\end{equation}
\begin{equation}
P_Y=\frac{\partial L}{\partial
\dot{Y}}=\frac{D}{2}\dot{X}+\frac{D(D-1)}{6}\dot{Y}, \label{34}
\end{equation}
from which we obtain
\begin{equation}
\dot{X}=\frac{6}{D+2}\left[P_X
\left(\frac{1-D}{3}\right)+P_Y\right], \label{34'}
\end{equation}
\begin{equation}
\dot{Y}=\frac{6}{D(D-1)}\left[P_Y\frac{2(1-D)}{D+2}-P_X\frac{D(1-D)}{D+2}\right].
\label{35}
\end{equation}
Substituting Eqs.(\ref{34'}), (\ref{35}) into the Hamiltonian
constraint (\ref{24}), we obtain
\begin{equation}
H=(1-D)P_X^2-\frac{6}{D}P_Y^2+6P_X P_Y=0. \label{36}
\end{equation}
Now, we may use the following quantum mechanical replacements
$$
P_X \rightarrow -i\frac{\partial}{\partial X}\:\:\:\:,\:\:\:\: P_Y
\rightarrow -i\frac{\partial}{\partial Y},
$$
by which the Wheeler-DeWitt equation is obtained
\begin{equation}
\left[(D-1)\frac{\partial^2}{\partial
X^2}+\frac{6}{D}\frac{\partial^2}{\partial
Y^2}-6\frac{\partial}{\partial X}\frac{\partial}{\partial
Y}\right]\Psi(X, Y)=0, \label{37}
\end{equation}
where $\Psi(X, Y)$ is the wave function of the universe in the
$(X, Y)$ mini-superspace.

We introduce the following change of variables
\begin{equation}
x=X(1-\frac{D}{D+3})+\frac{D}{D+3}Y \:\:\:\:,\:\:\:\:
y=\frac{X-Y}{D+3}, \label{38}
\end{equation}
by which the Wheeler-DeWitt equation takes a simple form
\begin{equation}
\left\{ -3\frac{\partial^2}{\partial x^2}+\frac{D+2}{D}
\frac{\partial^2}{\partial y^2}\right\} \Psi(x, y)=0. \label{39}
\end{equation}
Now, we can separate the variables as $\Psi(x, y)=\phi(x) \psi(y)$
to obtain the following equations
\begin{equation}
\frac{\partial^2 \phi(x)}{\partial x^2}=\frac{\gamma}{3} \phi(x),
\label{40}
\end{equation}
\begin{equation}
\frac{\partial^2 \psi(y)}{\partial y^2}=\frac{\gamma
D}{D+2}\psi(y), \label{41}
\end{equation}
where we assume $\gamma >0$.

\section{Solutions of Wheeler-DeWitt equation}

The solutions of Eqs.(\ref{40}), (\ref{41}) in terms of $x, y$ are
as follows
\begin{equation}
\phi(x)=e^{\pm\sqrt{\frac{\gamma}{3}}x}, \label{42}
\end{equation}
\begin{equation}
\psi(y)=e^{\pm\sqrt{\frac{\gamma D}{D+2}}y}, \label{43}
\end{equation}
leading to the four possible solutions for $\Psi(x, y)$ as
\begin{equation}
\Psi_D^{\pm} (x, y)= A^{\pm}e^{\pm\sqrt{\frac{\gamma}{3}}x \pm
\sqrt{\frac{\gamma D}{D+2}}y}, \label{43'}
\end{equation}
\begin{equation}
\Psi_D^{\pm} (x, y)= B^{\pm}e^{\pm\sqrt{\frac{\gamma}{3}}x \mp
\sqrt{\frac{\gamma D}{D+2}}y}, \label{44}
\end{equation}
or alternative solutions in terms of $X, Y$ as
\begin{equation}
\Psi_D^{\pm} (x, y)= A^{\pm}e^{\pm \sqrt{\frac{\gamma}{3}} \left(
\frac{3X+DY}{D+3}\right) \pm \sqrt{\frac{\gamma D}{D+2}}\left(
\frac{X-Y}{D+3}\right)}, \label{45}
\end{equation}
\begin{equation}
\Psi_D^{\pm} (x, y)= B^{\pm}e^{\pm \sqrt{\frac{\gamma}{3}} \left(
\frac{3X+DY}{D+3}\right) \mp \sqrt{\frac{\gamma
D}{D+2}}\left(\frac{X-Y}{D+3}\right)}, \label{46}
\end{equation}
where $A^{\pm}, B^{\pm}$ are the normalization constants. We may
also write down the solutions in terms of $R$ and $a$
\footnote{For $D=3$, there is a exchange symmetry $\Psi(R, a)
\leftrightarrow \Psi(a, R)$ under the exchange $ a \leftrightarrow
R$.}
\begin{equation}
\Psi_D^{\pm} (R, a)= A^{\pm} R^{\pm
\frac{1}{D+3}\left(\sqrt{3\gamma}+\sqrt{\frac{\gamma
D}{D+2}}\right)} a^{\pm \frac{1}{D+3} \left(
\sqrt{\frac{\gamma}{3}}D-\sqrt{\frac{\gamma D}{D+2}}\right)},
\label{47}
\end{equation}
\begin{equation}
\Psi_D^{\pm} (R, a)= B^{\pm} R^{\pm
\frac{1}{D+3}\left(\sqrt{3\gamma}-\sqrt{\frac{\gamma
D}{D+2}}\right)} a^{\pm \frac{1}{D+3}
\left(\sqrt{\frac{\gamma}{3}}D+\sqrt{\frac{\gamma
D}{D+2}}\right)}. \label{48}
\end{equation}

It is now important to impose the {\it good} boundary conditions
on the above solutions to single out the physical ones. In so
doing, we may impose the following condition
\begin{equation}
\Psi_D(R\rightarrow \infty, a\rightarrow \infty)=0,
\end{equation}
which requires the wave function of the universe to be
normalizable. This means that our minisuperspace model has no
classical solutions that expand simultaneously to infinite values
of $a$ and $R$, as Eqs.(\ref{30'})-(\ref{00}) show. Then, one may
take the following solutions
\begin{equation}
\Psi_D^{\pm} (R, a)= C^{\pm} R^{-
\frac{1}{D+3}\left(\sqrt{3\gamma}\pm\sqrt{\frac{\gamma
D}{D+2}}\right)} a^{-\frac{1}{D+3} \left(\sqrt{\frac{\gamma}{3}}D
\mp \sqrt{\frac{\gamma D}{D+2}}\right)}, \label{49}
\end{equation}
where $C^{\pm}$ are the normalization constants and the exponents
of $R$ and $a$ are negative for any value of $D$ \footnote{For
$D=1$, the exponent of ``$a$'' corresponding to $\Psi^+$ becomes
zero so that $\Psi^+$ depends only on $R$ with the condition
$\Psi^+(R \rightarrow \infty) \rightarrow 0$.}.

One may obtain the solutions (\ref{49}) in $(X, Y)$
mini-superspace as
\begin{equation}
\Psi_D^{\pm} (x, y)= C^{\pm}e^{-\sqrt{\frac{\gamma}{3}} \left(
\frac{3X+DY}{D+3}\right) \mp \sqrt{\frac{\gamma D}{D+2}}\left(
\frac{X-Y}{D+3}\right)}. \label{50}
\end{equation}

\section{Correspondence between classical and quantum cosmology}

One of the most interesting topics in the context of quantum
cosmology is the mechanisms through which the classical cosmology
may emerge from quantum theory. When does a Wheeler-DeWitt wave
function predict a classical space-time?  Quantum cosmology is the
quantum mechanics of an isolated system (universe). It is not
possible to use the Copenhagen interpretation, which needs the
existence of an external observer, since here the observer is part
of the system. Indeed, any attempt in constructing a viable
quantum gravity requires understanding the connections between
classical and quantum physics. Much work has been done in this
direction over the past decade. Actually, there is some tendency
towards using semiclassical approximations in dividing the
behaviour of the wave function into two types, oscillatory or
exponential which are supposed to correspond to classically
allowed or forbidden regions. Hartle \cite{5'} has put forward a
simple rule for applying quantum mechanics to a single system
(universe): {\it If the wave function is sufficiently peaked about
some region in the configuration space we predict to observe a
correlation between the observables which characterize this
region}. Halliwell \cite{6'} has shown that the oscillatory
semiclassical WKB wave function is peaked about a region of the
{\it minisuperspace} in which the correlation between the
coordinate and momentum holds good and stresses that both {\it
correlation} and {\it decoherence} are necessary before one can
say a system is classical. Using Wigner functions, Habib and
Laflamme \cite{7'} have studied the mutual compatibility of these
requirements and shown that some form of coarse graining is
necessary  for classical prediction from WKB wave functions.
Alternatively, Gaussian or coherent states with sharply peaked
wave functions are often used to obtain classical limits by
constructing wave packets.

In the investigation of classical limits, we first take $D=1$ and
look for a correspondence between classical and quantum solutions.
Using Eqs.(\ref{30'}) and (\ref{30''}) in the Planck units, the
corresponding classical locus in $(R, a)$ configuration space, is
\begin{equation}
Ra=1, \label{51}
\end{equation}
whereas in $(X, Y)$ coordinates we have
\begin{equation}
X+Y=0. \label{52}
\end{equation}
We now consider the wave functions (\ref{50}) in $(X, Y)$
mini-superspace for $D=1$
\begin{equation}
\Psi^+_1(X, Y)=C^+ e^{-\sqrt{\frac{\gamma}{3}}X}, \label{53}
\end{equation}
\begin{equation}
\Psi^-_1(X, Y)=C^- e^{-\sqrt{\frac{\gamma}{3}}\frac{X+Y}{2}}.
\label{54}
\end{equation}
The above wave functions, in their present form, are not square
integrable as is required for the wave functions to predict the
classical limit. However, one may take the absolute value of the
exponents to make the wave functions square integrable
\begin{equation}
\Psi^+_1(X, Y)=C^+ e^{-|\sqrt{\frac{\gamma}{3}}X|}, \label{55}
\end{equation}
\begin{equation}
\Psi^-_1(X, Y)=C^- e^{-|\sqrt{\frac{\gamma}{3}}\frac{X+Y}{2}|}.
\label{56}
\end{equation}
We next consider the general case $D>1$. Eliminating the parameter
$t$ in Eqs. (\ref{31}) and (\ref{32}) the classical loci in terms
of $R, a$ are obtained
\begin{equation}
a_\pm=R^{\frac{2}{D}\left[-1\pm
\sqrt{1-\frac{2}{3}(1-\frac{1}{D})}\right]^{-1} }. \label{57}
\end{equation}
The corresponding forms of these loci in terms of $X, Y$ are
\begin{equation}
Y_+=\frac{2}{D}X\left[-1+
\sqrt{1-\frac{2}{3}(1-\frac{1}{D})}\right]^{-1}, \label{58}
\end{equation}
\begin{equation}
Y_-=\frac{2}{D}X\left[-1-
\sqrt{1-\frac{2}{3}(1-\frac{1}{D})}\right]^{-1}. \label{59}
\end{equation}
The wave functions (\ref{50}) also are not square integrable, so
we may replace the exponents by their absolute values
\begin{equation}
\Psi_D^{\pm} (x, y)= C^{\pm}e^{-\left|\sqrt{\frac{\gamma}{3}}
\left( \frac{3X+DY}{D+3}\right) \mp \sqrt{\frac{\gamma
D}{D+2}}\left( \frac{X-Y}{D+3}\right)\right|}, \label{60}
\end{equation}
to make them square integrable. Now, following Hartle's point of
view, we try to make correspondence between the classical loci and
the wave functions.

Figures 1 - 6 show respectively the 2D plots of the typical wave
functions $\Psi^+_1$ - $\Psi^+_6$ in terms of $(X, Y)$ for
$\gamma=10^{-6}$; Figures 23 - 28 show the corresponding 3D plots,
respectively. On the other hand, Figures 12 - 17 show the
classical loci corresponding to $D=1 - 6$, respectively. It is
seen that the 2D and 3D plots of the wave functions $\Psi^+_1$ -
$\Psi^+_6$ are exactly peaked on the classical loci.

In the same way, Figures 7 - 11 show respectively the 2D plots of
the wave functions $\Psi^-_2$ - $\Psi^-_6$. Figures 29 - 33 show
the corresponding 3D plots, respectively. Figures 18 - 22 show the
classical loci for $D=2 - 6$, respectively. Again, an exact
correspondence is seen between the 2D and 3D plots of the wave
functions $\Psi^-_2$ - $\Psi^-_6$ and the classical loci. This
procedure will apply for all $D$.

\newpage
\section*{Concluding remarks}

First, we have studied a $(4+D)$-dimensional classical
Kaluza-Klein cosmology with a Robertson-Walker type metric having
two scale factors, $R$ for the universe and $a$ for the higher
dimensional space. By introducing a typical exotic matter with the
equation of state $p_\chi =(\frac{m}{3}-1)\rho_\chi$ in
4-dimensions, a decaying cosmological term is obtained effectively
as $\lambda \sim R^{-m}$. By taking $m=2$, the corresponding
Einstein field equations are obtained and we find exponential
solutions for $R$ and $a$ in terms of the Hubble parameter $H$.
These exponential solutions indicate the accelerating expansion of
the universe and dynamical compactification of extra dimensions,
respectively. It turns out that the rate of compactification of
extra dimensions as well as expansion of the universe depends on
the number of extra dimensions, $D$. The more extra dimensions,
the less rate of compactification and the more rate of
acceleration. It is worth noting that the model is free of initial
singularity problem because both $R$ and $a$ are non-zero at
$t=0$, resulting in a finite Ricci scalar.

Although the model describes in principle a closed universe with
non-vanishing cosmological constant, it is equivalent to a flat
universe with zero cosmological constant. Therefore, one may
assume that we are really living in a closed universe with
$\Lambda\neq 0$ , but it effectively appears as a flat universe
with $\Lambda= 0$. Note that we have not considered ordinary
matter sources in the model except an exotic matter source which
is to be considered as a source of dark energy. Therefore, it
seems the solutions to describe typical inflation rather than the
recently observed acceleration of the universe which is known to
take place in an ordinary matter dominated universe. However, if
the large percent of the matter sources in the universe would be
of dark energy type (as the present observations strongly
recommend), then one may keep the results here even in the
presence of other matter source, keeping in mind that the relevant
contribution to the total matter source of the universe is the
dark energy.

A question may arise on the fact that no physics is supposed to
exist below the planck length whereas for the contracting
solution, the scale factor $a(t)$ goes to zero starting from
$l_p$. However, it is not a major problem because we have not
considered elements of quantum gravity theory in this model and
merely studied a model based on general relativity which is
supposed to be valid in any scale without limitation. The scale
$l_p$, in this paper, is not introduced within a quantum gravity
model (action); it just appears as a typical initial condition, in
the middle of a classical model, based on the quantum cosmological
consideration. One may choose another scale based on some other
physical considerations.

We have also studied the corresponding quantum cosmology, through
the Wheeler-DeWitt equation, and obtained the exact solutions.
Based on Hartle'point of view on the correspondence between the
classical and quantum solutions, we have shown by 2D and 3D plots
of the wave functions a good correspondence between the classical
and quantum cosmological solutions for any $D$, provided that the
wave functions vanish for the infinite scale factors. There is no
such a correspondence if another boundary condition, other than
stated, is taken. Therefore, this correspondence guaranties that
the chosen boundary condition is a {\it good} one.

\newpage
{\large {\bf Figure captions}} \vspace{10mm}
\\
FIG. 1. 2D plot of $\Psi^+_1$ in terms of $(X, Y)$ for
$\gamma=10^{-6}$
\\
FIG. 2. 2D plot of $\Psi^+_2$ in terms of $(X, Y)$ for
$\gamma=10^{-6}$
\\
FIG. 3. 2D plot of $\Psi^+_3$ in terms of $(X, Y)$ for
$\gamma=10^{-6}$
\\
FIG. 4. 2D plot of $\Psi^+_4$ in terms of $(X, Y)$ for
$\gamma=10^{-6}$
\\
FIG. 5. 2D plot of $\Psi^+_5$ in terms of $(X, Y)$ for
$\gamma=10^{-6}$
\\
FIG. 6. 2D plot of $\Psi^+_6$ in terms of $(X, Y)$ for
$\gamma=10^{-6}$
\\
FIG. 7. 2D plot of $\Psi^-_2$ in terms of $(X, Y)$ for
$\gamma=10^{-6}$
\\
FIG. 8. 2D plot of $\Psi^-_3$ in terms of $(X, Y)$ for
$\gamma=10^{-6}$
\\
FIG. 9. 2D plot of $\Psi^-_4$ in terms of $(X, Y)$ for
$\gamma=10^{-6}$
\\
FIG. 10. 2D plot of $\Psi^-_5$ in terms of $(X, Y)$ for
$\gamma=10^{-6}$
\\
FIG. 11. 2D plot of $\Psi^-_6$ in terms of $(X, Y)$ for
$\gamma=10^{-6}$
\\
FIG. 12. Classical locus $X+Y=0$ for $D=1$
\\
FIG. 13. Classical locus $Y_-=\frac{2}{D}X[-1-
\sqrt{1-\frac{2}{3}(1-\frac{1}{D})}]^{-1}$ for $D=2$
\\
FIG. 14. Classical locus $Y_-=\frac{2}{D}X[-1-
\sqrt{1-\frac{2}{3}(1-\frac{1}{D})}]^{-1}$ for $D=3$
\\
FIG. 15. Classical locus $Y_-=\frac{2}{D}X[-1-
\sqrt{1-\frac{2}{3}(1-\frac{1}{D})}]^{-1}$ for $D=4$
\\
FIG. 16. Classical locus $Y_-=\frac{2}{D}X[-1-
\sqrt{1-\frac{2}{3}(1-\frac{1}{D})}]^{-1}$ for $D=5$
\\
FIG. 17. Classical locus $Y_-=\frac{2}{D}X[-1-
\sqrt{1-\frac{2}{3}(1-\frac{1}{D})}]^{-1}$ for $D=6$
\\
FIG. 18. Classical locus $Y_+=\frac{2}{D}X[-1+
\sqrt{1-\frac{2}{3}(1-\frac{1}{D})}]^{-1}$ for $D=2$
\\
FIG. 19. Classical locus $Y_+=\frac{2}{D}X[-1+
\sqrt{1-\frac{2}{3}(1-\frac{1}{D})}]^{-1}$ for $D=3$
\\
FIG. 20. Classical locus $Y_+=\frac{2}{D}X[-1+
\sqrt{1-\frac{2}{3}(1-\frac{1}{D})}]^{-1}$ for $D=4$
\\
FIG. 21. Classical locus $Y_+=\frac{2}{D}X[-1+
\sqrt{1-\frac{2}{3}(1-\frac{1}{D})}]^{-1}$ for $D=5$
\\
FIG. 22. Classical locus $Y_+=\frac{2}{D}X[-1+
\sqrt{1-\frac{2}{3}(1-\frac{1}{D})}]^{-1}$ for $D=6$
\\
FIG. 23. 3D plot of $\Psi^+_1$ in terms of $(X, Y)$ for
$\gamma=10^{-6}$
\\
FIG. 24. 3D plot of $\Psi^+_2$ in terms of $(X, Y)$ for
$\gamma=10^{-6}$
\\
FIG. 25. 3D plot of $\Psi^+_3$ in terms of $(X, Y)$ for
$\gamma=10^{-6}$
\\
FIG. 26. 3D plot of $\Psi^+_4$ in terms of $(X, Y)$ for
$\gamma=10^{-6}$
\\
FIG. 27. 3D plot of $\Psi^+_5$ in terms of $(X, Y)$ for
$\gamma=10^{-6}$
\\
FIG. 28. 3D plot of $\Psi^+_6$ in terms of $(X, Y)$ for
$\gamma=10^{-6}$
\\
FIG. 29. 3D plot of $\Psi^-_2$ in terms of $(X, Y)$ for
$\gamma=10^{-6}$
\\
FIG. 30. 3D plot of $\Psi^-_3$ in terms of $(X, Y)$ for
$\gamma=10^{-6}$
\\
FIG. 31. 3D plot of $\Psi^-_4$ in terms of $(X, Y)$ for
$\gamma=10^{-6}$
\\
FIG. 32. 3D plot of $\Psi^-_5$ in terms of $(X, Y)$ for
$\gamma=10^{-6}$
\\
FIG. 33. 3D plot of $\Psi^-_6$ in terms of $(X, Y)$ for
$\gamma=10^{-6}$

\newpage
\begin{thebibliography}{99}
\bibitem{paper}{\it This work is the extended version of the published paper,
Class.Quant.Grav. 20, 3385 (2003)}
\bibitem{1'}{\it S. Adler, Rev. Mod. Phys. {\bf 54}, 729 (1982); L. F.
Abbott, Phys. Lett. {\bf 150B}, 427 (1985); T. Banks, Nucl. Phys.
{\bf B249}, 332 (1985); S. M. Barr, Phys. Rev. D {\bf 36}, 1691
(1987); P. J. E. Peebles and B. Ratra, Astrophys. J. Lett. {\bf
325}, L17 (1988); B. Ratra and P. J. E. Peebles, Phys. Rev. D {\bf
37}, 3406 (1988).}
\bibitem{2'}{\it S. W. Hawking, Phys. Lett. {\bf 134B}, 403 (1984); E.
Baum, {\it ibid.}{\bf 133B}, 185 (1983); S. Coleman, Nucl. Phys.
{\bf B307}, 864 (1988); S. BGiddings and A. Strominger, Nucl.
Phys. {\bf B307}, 854 (1988); {\bf B321}, 481 (1989); T. Banks,
Nucl. Phys. {\bf B309}, 493 (1988).}
\bibitem{3'}{\it M. Ozer and M. O. Taha, Nucl. Phys. {\bf B287}, 776
(1987); K. Freese {\it et al, ibid}. {\bf B287}, 797 (1987); M.
Reuter and C. Wetterich, Phys. Lett. B {\bf 188}, 38 (1987); B.
Ratra and P. J. E. Peebles, Phys. Rev. D {\bf 37}, 3406 (1988); I.
Waga, Astrophys. J. {\bf 414}, 436 (1993); J. A. Frieman {\it et
al.}, Phys. Rev. Lett. {\bf 75}, 2077 (1995); K. Coble, S.
Dodelson, and J. A. Frieman, Phys. Rev. D {\bf 55}, 1851 (1997).}
\bibitem{1}{\it J. L. Lopez, D. V. Nanopoulos, Mod. Phys. Lett. A{\bf
11}, 1 (1996); M. \"{O}zer, M. O. Taha, Phys. Lett. B{\bf 171},
363 (1986); A.-M. M. Abdel-Rahman, Phys. Rev. D{\bf 45}, 3497
(1992); A. and R. G. Vishwakarma, Class. Quant. Grav. {\bf 14},
945 (1997); W. Chen, Y.-S. Wu, Phys. Rev. D{\bf 41}, 695 (1990);
M. O. Calvao {\em et al}, Phys. Rev. D{\bf 45}, 3869 (1992); V.
M\'{e}ndez, D. Pav\'{o}n, Gen. Rel. Grav. {\bf 28}, 697 (1996).}
\bibitem{2}{\it K. Freese {\em et al}, Nucl. Phys. B{\bf 287}, 797 (1987); M.
Gasperini, Phys. Lett. B{\bf 194}, 347 (1987); J. M. Overduin, P.
S. Wesson and S. Bowyer, Astrophys. J. {\bf 404}, 1 (1993).}
\bibitem{3}{\it F. Hoyle, G. Burbidge and J. V. Narlikar, Mon. Not. R. Astron.
Soc. {\bf 286}, 173 (1997).}
\bibitem{4}{\it M. A. Jafarizadeh, F. Darabi, A. Rezaei-Aghdam and A. R.
Rastegar, Phys. Rev. D{\bf 60}, 063514 (1999); T. S. Olson, T. F.
Jordan, Phys. Rev. D{\bf 35}, 3258 (1987); V. Silveira, I. Waga,
Phys. Rev. D{\bf 50}, 4890 (1994); Phys. Rev. D{\bf 56}, 4625
(1997); L. F. B. Torres, I. Waga, Mon. Not. R. Astron. Soc. {\bf
279}, 712 (1996).}
\bibitem{5}{\it M. D. Maia, G. S. Silva, Phys. Rev. D{\bf 50}, 7233 (1994); R.
F. Sister\'{o}, Gen. Rel. Grav. {\bf 23}, 1265 (1991); J.
Matyjasek, Phys. Rev. D{\bf 51}, 4154 (1995).}
\bibitem{6}{\it T. Appelquist, A. Chodos and P. G. O. Freund, {\it
Modern Kaluza-Klein Theories}, Frontiers in Physics Series,
(Volume {\bf 65}), 1986, (Ed. Addison-Wesely). A. G. Riess {\it et
al}. Astrophys. J. {\bf 560}, 49 (2001).}
\bibitem{A}{\it S. Perlmutter {\it et al}., Bull. Am. Phys. Soc. {\bf
29}, 1351 (1997); Ap. J. {\bf 507}, 46 (1998); A. G. Riess {\it et
al}., Astron. J. {\bf 116} (1998); P. M. Garnavich {\it et al}.,
Ap. J. Lett. {\bf 493}, 53 (1998); Science {\bf 279}, 1298 (1998);
Ap. J. {\bf 509}, 74 (1998); B. Schmidt {\it et al}., Astrophys.
J. {\bf 507}, 46 (1998).}
\bibitem{q}{\it L. Krauss and M. S. Turner, Gen. Rel. Grav. {\bf 27},
1137 (1995); J. P. Ostriker and P. J. Steinhardt, Nature {\bf
377}, 600 (1995); A. R. Liddle, D. H. Lyth, P. T. Viana and M.
White, Mon. Not. Roy. Astron. Soc. {\bf 282}, 281 (1996).}
\bibitem{!}{\it R. R. Caldwell, R. Dave and P. J. Steinhardt, Phys.
Rev. Lett. {\bf 80}, 1582 (1998); M. C. Bento, O. Bertolami, Gen.
Rel. Grav. {\bf 31}, 1461 (1999); O. Bertolami, Nuovo Cimento,
B{\bf 93}, 36 (1986).}
\bibitem{!!}{\it C. L. Benett {\it et al}., Astrophys. J. Supll. {\bf 148}, 1
(2003).}
\bibitem{!!!}{\it G. W. Gibbons, {\it Aspects of Supergravity Theories
in Supersymmetry, Supergravity and Related Topics}, (World
Scientific Singapore, 1985); J. Maldacena, C. Nu\~{n}ez, Int. J.
Mod. Phys. A{\bf 16}, 822 (2001).}
\bibitem{&&}{\it P. K. Townsend, M. N. R. Wohlfarth, Phys. Rev. Lett. {\bf 91},
061302 (2003).}
\bibitem{&&&}{\it L. Cornalba, M. Costa, Phys. Rev. D{\bf 66}, 066001
(2002); L. Cornalba, M. Costa, and C. Kounnas, Nucl. Phys. B{637},
378 (2002); N. Ohta, Phys. Rev. Lett. {\bf 91}, 061303 (2003); S.
Roy, Phys. Lett. {\bf B568}, 1 (2003).}
\bibitem{7}{\it A. Chodos and S. Detweiler, Phys. Rev. D{\bf 21}, 2167
(1980); T. Dereli and R. W. Tucker, Phys. Lett. {\bf B125}, 133
(1983); J. Gu, W. P. Hwang, Phys. Rev. D{\bf 66}, 024003 (2002).}
\bibitem{8}{\it A. S. Majumdar and S. K. Sethi, Phys. Rev. D{\bf 46}, 5315 (1992);
A. S. Majumdar, T. R. Seshadri and S. K. Sethi, Phys. Lett. B{\bf
312}, 67 (1993); A. S. Majumdar, Phys. Rev. D{\bf 55}, 6092
(1997); A. S. Majumdar, Ind. Jour. Phys.{\bf 73} B, 843 (1999); A.
S. Majumdar, Phys. Rev. D{\bf 64}, 083503 (2001).}
\bibitem{9}{\it N. Mohammedi, Phys. Rev. D{\bf 65}, 104018
(2002).}
\bibitem{@@}{\it D. Sahdev, Phys. Lett. B{\bf 137} 155 (1984).}
\bibitem{10}{\it J. Wudka, Phys. Rev. D{\bf 35}, 3255 (1987); Phys.
Rev. D{\bf 36}, 1036 (1987).}
\bibitem{!&}{\it G. Dvali, G. Gabadadze, M. Porrati, Phys. Lett. B{\bf
485}, 208 (2000); B. Abdesselam, N. Mohemmedi, Phys. Rev. D{\b65},
084018 (2002); N. Kaloper, J. March-Russell, G. D. Starkman and M.
Trodden, Phys. Rev. Lett. {\bf 85}, 928 (2000).}
\bibitem{12}{\it D. Atkatz and H. Pagels,
Phys. Rev. D{\bf 25}, 2065 (1982); Am. J. Phys. {\bf 62} (7), 619
(1994).}
\bibitem{15}{\it M. A. Jafarizadeh, F. Darabi, A. Rezaei-Aghdam and A.
R. Rastegar, Mod.Phys.Lett. A{\bf 13}, 3213 (1998).}
\bibitem{*}{\it P. J. E. Peebles, B. Ratra, Rev. Mod. Phys. {\bf 75}, 559
(2003).}
\bibitem{13}{\it W. Chen and Y. Wu, Phys. Rev. D{\bf 41}, 695
(1990).}
\bibitem{4'}{\it B. S. DeWitt, Phys. Rev. {\bf 160}, 1113 (1967).}
\bibitem{Dirac}{\it P. M. A. Dirac, Lectures on Quantum Mechanics,
Yeshiva University, (Academic press, New York) 1967.}
\bibitem{5'}{\it J. B. Hartle, in Gravitation in Astrophysics, 1986 NATO
Advanced Summer Institude, Garg\`{e}se, edited by B. Carter and J.
Hartle  (NATO ASI Series B: Physics Vol. {\bf 156}, Plenum. New
York) 1987.}
\bibitem{6'}{\it J. J. Halliwell, Phys. Rev. D {\bf 39} 2912
(1989).}
\bibitem{7'}{\it S. Habib, R. Laflamme, Phys. Rev. D {\bf 42} 4056 (1990);
S. Habib, Phys. Rev. D {\bf 42} 2566 (1990).}
\end{thebibliography}
\end{document}